\documentclass[11pt]{article}

\newcommand{\be}{\begin{equation}}
\newcommand{\ee}{\end{equation}}
\newcommand{\Ref}[1]{(\ref{#1})}
\def \ni{\noindent}
\def \mn{\medskip
\ni}
\def \f{\frac}
\def \R{{\bf R}}
\let \pp=\partial
\def \trn{{Tr}_{N\times N}}
\def \trg{{Tr}_{\cal G}}
\def \tl{\tilde}
\newcommand{\mat}[2]{\left(\begin{array}{#1}#2
\end{array}\right)} 
\def \qp{Q_{+1}}
\def \qm{Q_{-1}}

\begin{document}

\title{BRST quantization of Matrix Chern-Simons Theory}

\author{{\bf Etera R. Livine}\thanks{e.mail: livine@cpt.univ-mrs.fr} \\
Centre de Physique Th{\'e}orique,
Campus de Luminy, Case 907, \\
13288 Marseille cedex 9, France \vspace{2mm}\\
{\bf Lee Smolin}\thanks{e.mail:lsmolin@perimeterinstitute.ca}\\
Perimeter Institute for Theoretical Physics, \\
Waterloo, Ontario, Canada N2J 2W9 \\
and \\
Department of Physics, University of Waterloo}
\date{\today}
\maketitle

\begin{abstract}
The $BRST$ quantization of matrix Chern-Simons theory is carried out,
the symmetries of the theory are analyzed and used to constrain the form of the effective action. 

\end{abstract}


\newpage

\section{Introduction}

The cubic matrix models\cite{cubic,stringloop,kawaietal,student} were 
invented as a possible approach to the problem of formulating
string or $\cal M$ theory in a background invariant framework. They
also provide a matrix formulation of certain quantum
deformed extensions of loop quantum
gravity\cite{stringloop,tubes,pqtubes}.  
The basic philosophy that motivates these theories is that
quantum and classical theories of gravity which are background
independent can, in most if not all of the known cases, be 
constructed by modifying topological field theories.  The idea
is then to construct matrix models that extend a 
matrix form of 
Chern-Simons theory\cite{cubic,stringloop}.

The quantization of these theories faces certain issues due to the
fact that the action is presented in a first order form, which means 
that they define theories on phase spaces rather than configuration 
spaces. These theories also have gauge symmetries and constraints that
must be taken into account correctly.  In this letter we present
an approach to quantization based on the standard $BRST$ method.
We carry out the quantization in detail for the case of
Matrix Chern-Simons theory, with and without couplings to fermions,  
and show that it leads to results which
are direct extensions of the usual results for Chern-Simons theory.
These results should be directly extendable to the cubic matrix 
models.

While these results are encouraging, we must, 
however, mention another important issue that we do not 
solve in this paper. This is to give a genuinely background 
independent quantization of the cubic matrix models.  What we describe here
is instead a {\it background dependent quantization} of a theory whose classical
formulation is background independent. This is because in a 
$BRST$ formalism the quantum 
theory is defined {\it relative to a given background}, which is a
solution to the classical equations of motion of the theory. Whether 
there is a form of the quantum theory that is well defined at the more 
fundamental, background independent, level, which unifies the 
particular background dependent quantum theories, remains an open 
question\footnote{One of us has explored an approach to a background
independent quantum theory based on a hidden variable theory, and 
stochastic quantization\cite{hiddenmatrices}.  Other authors 
have also noted that matrix models are in some sense automatically
hidden variables theories, in that quantum theory can be defined from
their classical statistical mechanics\cite{adler,artem}.  Whether 
there is a less radical approach to background independent 
quantization of matrix models is an open question.}. 

In the next section we review Matrix Chern-Simons theory. We
discuss the symmetries of the theory and give 
several examples including a possible matrix version of $2+1$ gravity 
and
a supersymmetric model. The main technical work of the paper is in 
section 3, where we discuss gauge fixing, ghosts, the $BRST$ and 
anti-$BRST$ transformations.  In section 4  we show that pure Matrix 
Chern-Simons
theory has also a vector supersymmetry, and use this to discuss the
form of the effective action, with and without fermions.

\section{Matrix Chern-Simons theory}

\subsection{The Cubic matrix model}

We consider an action $S[A]$ where $A=A_a\tau^a$,
the $A_a$ being $N\times N$  matrices
and the $\tau^a$ the generators of a Lie algebra ${\cal G}$
in a given representation.
We can also use a superLie algebra, in which case
we need to use the supertrace
instead of the trace and the supercommutators between elements of the
 ${\cal G}$ algebra.
We will use the indices $i,j,k$ for the $N\times N$ indices
and $\alpha,\beta,\gamma$ for the ${\cal G}$ representation.
We introduce the cubic matrix model:

\be
S[A]=\trn A_\alpha^\beta[A_\beta^\gamma,A_\gamma^\alpha]=
\trg A_i^j[A_j^k,A_k^i]
\label{S}
\ee
where $\trg$ is the trace for the ${\cal G}$ representation and
$\trn$ the trace for $N\times N$ matrices. The action
can also be written as

\be
S[A]=\left(\trg \tau^a\tau^b\tau^c\right)
\left(\trn A_a[A_b,A_c]\right)
=\f{1}{3}\varphi^{abc}\trn A_a[A_b,A_c]
\ee

where $\varphi^{abc}= 3/2\trg \tau^a [\tau^b,\tau^c]$ is the
usual structure constant $f^{bc}{}_a$ 
with the third index raised by the metric $\eta^{ab}={Tr}(\tau^a\tau^b)$:
\, $\varphi^{abc}=\eta^{ad}f^{bc}{}_d$

This cubic action $S$ has two global symmetries.
First, it is invariant under the global action of the group $G$
generated by ${\cal G}$:

\be
A_i^j\rightarrow R^{-1}(g)A_i^jR(g) 
\label{G}
\ee
where $R(g)$ is the representation matrix of the group element 
$g\in G$. $g$ acts on the generators $\tau^a$ by conjugation.
The infinitesimal version of this symmetry is

\be
\delta A_i^j = \epsilon [A_i^j,u] \textrm{ with } u\in {\cal G}
\ee

The second symmetry is an invariance under rotation by the group
$G_N=GL_N(\R)$:

\be
A_\alpha^\beta\rightarrow M^{-1}A_\alpha^\beta M
\textrm{ or equivalently }
A_a \rightarrow M^{-1}A_aM
\label{gaugeS}
\ee
for $M\in G_N$. The infinitesimal variation is then given
by a $N\times N$ matrix $m$:
\be
\delta A_a=\epsilon[A_a,m] 
\ee

\medskip
Next, we can look at the classical solutions $X$ of the action $S$.
They are given by the equation
\be
\varphi^{abc}[X_b,X_c]=0
\ee
Let point out that the set of solutions is invariant under both
$G$ and $G_N$ rotations.
We can study the fluctuations of our matrix around the new background
given by $X$
by introducing the new action

\be
S_X[A]=S[X+A]-S[X]=
\varphi^{abc}
\left(\trn A_a[X_b,A_c] +\f{1}{3}\trn A_a[A_b,A_c]\right)
\label{SX}
\ee
The $G_N$ symmetry now reads

\be
M^{-1}(X_a+A_a)M=
M^{-1}X_aM+M^{-1}A_aM=X_a+A^{(M)}_a
\ee
so that $S_X$ has a $G_N$ gauge symmetry given by

\be
A_a\rightarrow A^{(M)}_a=
M^{-1}A_aM+M^{-1}[X_a,M]_N
\label{gaugeSX}
\ee
which shows that the background $X_a$ takes the role of a derivation,
similarly to the differential calculus in Non-Commutative Geometry.
In this setting, $A_a$ behaves like $G_N$ gauge field. Let nevertheless
point out that if $X_a=0$ then it is not a gauge field anymore but behaves
simply like a $G_N$ matter field.

The action $S_X$ also has a $G$ gauge symmetry given by
\be
A_i^j\rightarrow A_i^j{}^{(g)}=
g^{-1}A_i^jg+g^{-1}[X_i^j,g]_{\cal G}
\ee
Here too, $A_i$ behaves like a $G$ gauge field when $X_i^j\ne 0$ and
like a matter field otherwise. An interesting configuration is when
$(X_a)_i^j$ is a diagonal $N\times N$ matrix, so that the diagonal
fields $A_i^i$ are gauge fields and the off-diagonal elements $A_i^j$
are the matter fields. 

\subsection{Examples}

Matrix Chern-Simons theory is the particular
case when we choose $G=SU(2)$. Then the generators are

\be
\tau_1=\f{1}{\sqrt 2}\mat{cc}{0 & 1 \\ 1 & 0} \quad
\tau_2=\f{1}{\sqrt 2}\mat{cc}{0 & i \\ -i & 0} \quad
\tau_3=\f{1}{\sqrt 2}\mat{cc}{1 & 0 \\ 0 & -1}
\ee
Then the structure constants are given by the antisymmetric tensor
$\varphi^{abc}=i\epsilon^{abc}$ and the metric
$\eta^{ab}=\delta^{ab}$ is trivial. The classical solutions
are sets of three matrices $X_1,X_2,X_3$ which commute with each other.
Then the action $S_X$ looks like a discrete version of the
usual Chern-Simons action. More precisely, after a triple
compactification \cite{cubic} achieved through a special choice
of background solutions $X$ in a large matrix limit, the trace
reproduces the integration over a 3\_torus and the matrix Chern-Simons
action exactly reproduces the usual Chern-Simons field theory.

Another similar example is given by the choice $G=SL(2,R)$. Then the generators
would be

\be
\tau_1=\f{1}{\sqrt 2}\mat{cc}{0 & 1 \\ 1 & 0} \quad
\tau_2=\f{1}{\sqrt 2}\mat{cc}{0 & 1 \\ -1 & 0} \quad
\tau_3=\f{1}{\sqrt 2}\mat{cc}{1 & 0 \\ 0 & -1}
\ee
Then the structure constants are once again given by the antisymmetric tensor
$\varphi^{abc}=\epsilon^{abc}$ but
the metric now has a $(-,+,+)$ signature instead of the previous $(+,+,+)$.

One can also investigate a supersymmetric extension of $SU(2)$ by
considering the superalgebra ${\cal G}=osp(1|2)$. This introduces
odd-Grassmann generators and a spinor degree of freedom.
This adds the fermionic term:

\be
S^{fermions}=\trn \phi_B [A_a,\phi_A] (\tau^a)_A{}^{-B}
\ee
where $\phi$ is the spinor and $A,B=\pm 1/2$ the spinor indices.
We can choose
the classical solution to have components only in the $su(2)$ generators
so that we get fermionic degrees of freedom behaving like matter
fields.
Let us point out that this is not the classical supersymmetric
extension
of the Chern-Simons action. Indeed, the usual supersymmetry is coupled
to the Poincare group whereas, in our case, it is an extension of the
Lorentz symmetry: we couple it to the frame rotations and not to the
translations.  

We can also turn to groups larger than $SU(2)$.  An interesting example
is given by $SL(2,C)$, which is the complexification of $SU(2)$. Indeed its
generators are the $J^a=\tau^a$ and the $K^a=i\tau^a$. Let's take
$A=A_a\tau^a+iE_a\tau^a$ and write the cubic action
choosing the fundamental $2\times 2$ representation of $SL(2,C)$:

\be
S[A]=-\epsilon^{abc}\left(\trn(E_a[A_b,A_c])
+\trn(E_a[E_b,E_c])\right)
+i\epsilon^{abc}\left(\trn(E_a[A_b,E_c])
+\trn(A_a[A_b,A_c])
\right)
\ee
The first term here looks very much like the $EF+EEE$ action of $2+1$d
gravity with cosmological constant (obtained by rescaling the matrix
$E$).
Indeed after (triple) compactification, we indeed find back exactly
that action. 
The $i$ term is the extra term coming in
Witten's reformulation of $2+1$d gravity as a $SL(2,C)$ Chern-Simons theory
\cite{witten}.

\subsection{Symmetries and the physical Interpretation}

We can notice that in addition to the global gauge symmetries, there
are local versions of the gauge symmetries
\Ref{G} and \Ref{gaugeS}. The action $S$ is further invariant under
the transformations

\be
A_i^j\rightarrow g_i^{-1}A_i^jg_j
\label{localG}
\ee
and
\be
A_\alpha^\beta\rightarrow M_\alpha^{-1}A_\alpha^\beta M_\beta
\label{localN}
\ee

One has a nice interpretation of the local gauge symmetry \Ref{localG}
in the context of M-theory \cite{taylor}. The matrix $(A_i^j)_{1\le i,j\le N}$
represents the interactions (due to open strings) between $N$ D0-branes (or
equivalently $N$ points). And one has a local gauge symmetry $G$ at 
each of these points,
which gives \Ref{localG}.

One also has ``translation'' symmetries:

\be
A_a\rightarrow A_a+\lambda_a Id_N
\qquad
A_i^j\rightarrow A_i^j+\lambda_i^j Id_G
\label{translation}
\ee

The link between the matrix models and the usual physical actions 
goes usually through compactification procedures \cite{taylor}
which creates dimensions and a space-time out the matrix in the limit
$N\rightarrow\infty$. Then, one finds back for example the usual
Chern-Simons theory on a 3d manifold out of the Matrix CS model described above
\cite{cubic,stringloop}. In this context, the translation symmetries \Ref{translation}
really become the symmetry by translation in the emerging space-time.

Nevertheless, it would be interesting to give a meaning to the matrix models
for $N$ finite without talking about the possible infinite matrix limit.
In sight of the expression \Ref{SX}, one automatically thinks about a potential link
with non-commutative geometry (see \cite{landi} for example) with $X$
being the Dirac operator governing differential calculus. However,
the $A=A_a\tau^a$ do not form an algebra whose product could help us construct
the actions $S$ or $S_X$. Still, there is some hope in using the algebra
of $2\times 2$ matrices to translate the matrix model into the spectral triple
language. We could then make the gauge group $SU(2)$ appear as
the unitary part of ${\cal M}_2(C)$. Or we could say that using the algebra
${\cal M}_2(R)$ in the cubic matrix model defined is equivalent
to using the algebra $sp(2)$.
These possibilities will be investigated in future work.

\section{Gauge fixing the cubic action}


One interesting and necessary step in studying the actions \Ref{S} or
\Ref{SX} is to gauge fix them. In our case, we study the gauge fixing
of the $G_N$ group since it is the apparent gauge symmetry of the
action \Ref{SX}. Nevertheless, the same techniques work perfectly for the
gauge fixing of the $G$ symmetry.

The first gauge fixing procedure which comes at one's mind is choosing
a representant for each orbit under the action by $G_N$ conjugation
\footnote{One can carry out the BRST analysis in that
case the same way as in the case of the Landau gauge which we
present. Let us choose a section $s$ of the orbits of $\theta^aA_a$,
where $\theta^a$ is a fixed vector.
Then the gauged fixed action is
$$
\tl{S}=
-\f{1}{6}Tr(A_a[A_b,A_c])+Tr (W(\theta^aA_a-s(\theta^aA_a)))+
Tr (U[\theta^aA_a,V]]).
$$
The residual BRST symmetry is the same as in equation \Ref{brstsym}
and ensures invariance of the path integral under change of section.}.
However, a more physically interesting choice is the Landau gauge,
in which we find back the usual features of the analysis of
the Chern-Simons action.

\subsection{Landau gauge}

We choose a discrete equivalent of the Landau gauge $\pp_aA^a=0$
as gauge fixing condition:

\be
\trg [X,A]_N =[X_a,A^a]_N=0
\textrm{ where the metric is } \eta^{ab}={Tr}(\tau^a\tau^b)
\ee

\mn
Let first look at the gauge fixing of the initial action $S$.
To calculate the ghost term arising from the break
of the $G_N$ symmetry, we calculate
the variation of $[X,A]$ under a small gauge transformation
\Ref{gaugeS}
given by $M=e^m =1+m + \dots$:

\be
\delta_m [X_a,A^a]=[X_a,[A^a,m]]
\ee
so that the ghost term introduces two odd-Grassmann
valued $N\times N$ matrices $U$ and $V$:

\be
\trn (U[X_a,[A^a,V]]_N)
=\trn \trg (U[X,[A,V]_N]_N)
\ee
Now the entire gauge fixed part action reads

\be
\tl{S}=S+S^{ghost}=
-\f{1}{2}\times\f{1}{3}Tr(A_a[A_b,A_c])+Tr(W[X_a,A^a])+
Tr(U[X_a,[A^a,V]])
\label{totalS}
\ee
where the even-Grassmann $N\times N$ matrix
$W$ enforces the gauge fixing condition. Through this procedure,
we can introduce a background $X$ in the background independent
action $S$.

The gauge fixing of the action $S_X$ is very similar to the one of $S$.
Indeed, the variation of the gauge fixing condition is now
\be
\delta_m [X_a,A^a]=[X_a,[X^a+A^a,m]]
\ee
so that it is sufficient to replace $A$ by $X+A$ in the previous
calculations:
\be
S^{ghost}_X[A,W,U,V]
=\trn \left(W[X^a,A_a]\right)+\trn\left(U[X^a,[X_a+A_a,V]]\right)
\ee
and
\be
\tl{S}_X=S_X+S^{ghost}_X=
-\f{1}{2}\epsilon^{abc}\left(Tr(A_a[X_b,A_c])
+\f{1}{3}Tr(A_a[A_b,A_c])\right)+
Tr(W[X_a,A^a])+
Tr(U[X_a,[X^a+A^a,V]])
\label{totalSX}
\ee

The resulting action $\tl{S}_X$ has the exact same structure
as the gauge-fixed Chern-Simons action \cite{delduc} and we can similarly
find the BRST transformations under which $\tl{S}_X$ is invariant.
In the following paragraph,
we are going to write down the BRST generators in the case of
$\tl{S}$, keeping in mind that they can be
easily generalized to $\tl{S}_X$ by changing $A$ into $X+A$ in the different
formulas.

\subsection{BRST transformations}

\mn
The gauge fixed action $\tl{S}$
is invariant under the following BRST transformations
where $\epsilon$ is a odd-Grassmann valued number:

\begin{equation}
\left\{
\begin{array}{ccc}
\delta A_a &=  &[A_a,V]\epsilon \\
\delta U & =& W\epsilon \\
\delta V & =& V^2 \epsilon \\
\delta W & =& 0
\end{array}
\right.
\label{brstsym}
\end{equation}
It is simply a gauge transformation for the initial action $S$
and it is easy to check the ghost part $S_{ghost}$ is
also invariant under these transformations.
Thus, we can introduce the BRST charge $\qp$ acting as:

\begin{equation}
\left\{
\begin{array}{ccc}
\qp A_a &=  & -[A_a,V] \\
\qp U & =& W\\
\qp V & =& V^2  \\
\qp W & =& 0
\end{array}
\right.
\end{equation}
We can then rewrite the ghost part of the action as
\begin{equation}
S^{ghost}=\trn\left(-U[X^a,\qp A_a]\right)-\trn\left(A_a[X^a,\qp U]\right)
\end{equation}
from which it is straightforward to check the invariance under $\qp$.

\medskip

We can also write down the action as if we had done an integration by parts
on the ghost term. Then, the gauge fixing part of the action reads:

\begin{equation}
S^{ghost}=\trn \left (
(W-\{U,V\})[X^a,A_a]
\right )-
\trn \left (
V[X^a,[A_a,U]]
\right )
\end{equation}

\ni
The action written in this form has a similar BRST invariance as above.
It is generated by what we call the anti-BRST operator $\qm$:

\be
\left\{
\begin{array}{ccc}
\qm A_a &=  & -[A_a,U] \\
\qm U & =& U^2\\
\qm V & =& -W+\{U,V\}  \\
\qm W & =& [U,W]
\end{array}
\right.
\ee
We have the following commutation relations between
the BRST charges:

\be
\qp^2=0 \quad \qm^2=0 \quad \{\qp,\qm\}=0
\ee
Moreover, we can assign the ghost number 0 to $A$ and $W$, -1 to $U$ and
+1 to $V$. Then, $\qp$ increases the ghost number by an unit, and $\qm$
decreases it by an unit.

\section{The effective action}

Once we have gauged fixed the action, one would like to compute its loop
expansion and the resulting effective action. In this section, we will restrict
ourself to the study of the Matrix Chern-Simons model $G=SU(2)$.
One can not apply the usual techniques of perturbative expansion
for the gauge fixed action $\tl{S}$ for it doesn't have any quadratic term.
On the other hand, one can use the action $\tl{S}_X$ since the background $X$ introduces
propagators for the matrix $A$ and the ghost matrices $U,V$.
Moreover, the introduction of the matrix $W$ allows to invert the quadratic terms in order
to derive the propagator of the $A$ matrices. More precisely,
let's note $x_a=[X_a,.]$. As the $X_a$ matrices commute, the morphisms
$x_a$ also commute. Then, the matrix correlating the $A$ and $W$ is:

\be
C=\left(
\begin{array}{ccc|c}
0& x_3 & -x_2 & x_1 \\
-x_3 & 0 &x_1 & x_2 \\
x_2 & -x_1 & 0 & x_3 \\
\hline
-x_1 & -x_2 & -x_3 & 0
\end{array}
\right)
\ee
whose inverse is the propagator $P=C^{-1}=-RC$ where
we have introduce the matrix $R=(x_1^2+x_2^2+x_3^2)^{-1}$.
The propagator of the ghost is simply the matrix $R$ and
we have two types of 3-vertices $AAA$ and $UVA$. Then, one can easily
check by hand that the only 1-loop and 2-loop corrections are only of the type
$A[A,A]$ and that all other possible terms are canceled.
This comes from an additional symmetry of the matrix Chern-Simons action,
similar to the  so-called vector supersymmetry (VSUSY)
of the ordinary Chern-Simons theory.
This symmetry is special to the case $G=SU(2)$ (and also $G=SU(1,1)$) 
for which $\varphi^{abc}=\epsilon^{abc}$ ($a,b,c=1,2,3$).
In the continum limit
$N\rightarrow\infty$ in which we recover the full Chern-Simons theory,
it protects the theory from
infrared effects and contributes to the finiteness (or disappearing
depending of the regularization scheme) of the
quantum corrections.  The symmetry for $\tl{S}_X$ \Ref{totalSX}
reads with $\alpha=1,2,3$:

\be
\left\{\begin{array}{ccc}
\delta_\alpha A_a &=  &\epsilon_{a\alpha b}[X_b,V] \epsilon_\alpha\\
\delta_\alpha U & =& A_\alpha \epsilon_\alpha\\
\delta_\alpha V & =& 0 \\
\delta_\alpha W & =& [X_\alpha+A_\alpha,V] \epsilon_\alpha
\end{array}\right.
\quad \epsilon_\alpha\in R
\ee
In fact, this supersymmetry also exists for $\tl{S}$ \Ref{totalS} and reads:

\be
\left\{\begin{array}{ccc}
\delta_\alpha A_a &=  &\epsilon_{a\alpha b}[X_b,V] \epsilon_\alpha\\
\delta_\alpha U & =& A_\alpha \epsilon_\alpha\\
\delta_\alpha V & =& 0 \\
\delta_\alpha W & =& [A_\alpha,V] \epsilon_\alpha
\end{array}\right.
\quad \epsilon_\alpha\in R
\ee

It is not difficult to see that, as in ordinary Chern-Simons theory, 
the only term that can appear in the effective action 
which is invariant under local gauge, $BRST$ and vector supersymmetry 
transformations
is the original action itself.  The result can then only be a correction
in the coupling $k$. 

We may also consider the inclusion of 
fermions,  through a term such as
\begin{equation}
I^{\Psi} = Tr \left (  \Psi^A  [X_a , \Psi^B ]   \right ) \tau^{a}_{AB}.
\end{equation}
Such a term can be introduced by considering the cubic matrix model associated
to the superalgebra $osp(1|2)$.
It is not hard to see that it breaks the vector supersymmetry.
Then, the theory also knows about a background metric, formed by
$q^{ab}=Tr \tau^a \tau^b$.  
The remaining $BRST$ invariance then allows the appearance of a 
Yang-Mills like term in the effective action, of the form,
\begin{equation}
S_{1 loop} = c Tr \left (  [X_a , X_b][X_c , X_d]   \right ) q^{ac}q^{bd}
\end{equation}

So long as $N$ is finite this term need not be considered part of the 
fundamental action, but only as a part of the effective theory 
governing low energy phenomena. But if we take $N \rightarrow \infty$
then it may become necessary to introduce this as a fundamental
term in the action, as in continuum Chern-Simons-Yang-Mills
theory\cite{topologicalmassive}.

\section{Conclusions}

The results of this, fairly straightforward technical paper,
are interesting first of all for the project of basing
$\cal M$ theory on a cubic matrix model; we see that the
$BRST$ quantization does suffice to base a perturbative quantization
of these theories around backgrounds defined by classical solutions
to their field equations.  Moreover the fact that, as expected, the
results mirror those of ordinary Chern-Simons theory, with and
without coupling to fermions confirms the physical picture
behind the loop/string duality postulated in \cite{stringloop}.
The basic physical idea there is that, as in the case of topologically massive gauge theories in $2+1$ dimensions, 
these theories will have
two phases, one background dependent and one background independent. 
In the latter  the degrees of freedom on the toroidal 
compactifications will be Chern-Simons like, which means that the
perturbation theory of the matrix models is independent of the metric
structure defined by the toroidal compactification. As it results
it generates an extension
of loop quantum gravity of the kind described in \cite{tubes,pqtubes}. 
However, in the other phase,  the degrees of freedom
include (topologically) massive quanta of $3d$ Yang-Mills theories
which, in the case of compactifications of the kind described
in \cite{cubic,stringloop} become modes of strings\footnote{We note that
this is similar to the physical picture of Kogan et al\cite{Kogan}
but realized precisely in the context of a matrix model.}.
The possibility that both behaviors can arise as different phases
of a single matrix theory is the dynamical basis of the conjecture
that loop quantum gravity, at least with certain choices
of representation labels, and string theory, may be dual descriptions
of the same theory.

\section*{Acknowledgements}

We would like to thank Yi Ling and Fotini Markopoulou for
conversations during the course of this work. 
We are very grateful for support by the NSF through grant
PHY95-14240 and gifts from the Jesse
Phillips Foundation.

\end{document}